%% file: main.tex
\documentclass[10pt]{IEEEtran}

\usepackage{amssymb,amsthm,mathrsfs,mathtools,amsfonts}
\usepackage[normalem]{ulem}
\usepackage{bm}

\usepackage{framed}
\usepackage{wrapfig}
\usepackage{algorithmic}
\usepackage{algorithm}
\algsetup{linenosize=\scriptsize}
\usepackage{amsmath}
\usepackage[dvipsnames]{xcolor}
\usepackage{cite}
\usepackage{url}
\usepackage[final]{graphicx}
\usepackage{subfigure}
\graphicspath{{./}{./figures/}}
\renewcommand{\figurename}{Fig.}

\input{commands.tex}


\title{Deep Neural Network Feature Designs for RF Data-Driven Wireless Device Classification}

\author{Bechir Hamdaoui, Abdurrahman Elmaghbub, Seifeddine Mejri
\thanks{This research is supported in part by the National Science Foundation under grant award No. 2003273.}
\thanks{An IEEE-formatted version of this article is published in IEEE Network. Personal use of this material is permitted. Permission from IEEE must be obtained for all other uses, in any current or future media, including reprinting/republishing this material for advertising or promotional purposes, creating new collective works, for resale or redistribution to servers or lists, or reuse of any copyrighted component of this work in other works.}
~\\
\small Oregon State University, \small Corvallis, Oregon~\\
\small \{hamdaoui,elmaghba,mejris\}@oregonstate.edu~\\
}

\IEEEoverridecommandlockouts
\twocolumn

\begin{document}

\maketitle

\begin{abstract}
\input{abstract.tex}
\end{abstract}


\section{Introduction}
\input{introduction.tex}

\section{RF Data-Driven DNN Features: Current Limitations and Design Goals}
\label{sec:limiations-features}
\input{prior-feature-limitation.tex}

\section{Feature Design Approach I: Exploiting Out-of-Band Distortion Information}
\label{sec:hw-features}
\input{hw-features.tex}

\section{Feature Design Approach II: Exploiting RF Spectrum Domain Knowledge}
\label{sec:domain}
\input{domain-features.tex}

\section{Other Feature Design Ideas and Their Unsolved Research Challenges}
\label{sec:open-challenges}
\input{open-challenges.tex}

\section{Conclusion}
\input{conclusion.tex}

\bibliographystyle{IEEEtran}

\newpage
\noindent {\footnotesize {\bf Bechir Hamdaoui} (S'02-M'05-SM'12) is a Professor in the School of EECS at Oregon State University. He received M.S. degrees in both ECE (2002) and CS (2004), and the Ph.D. degree in ECE (2005) all from the University of Wisconsin-Madison. His research interests are in the general areas of computer networks, wireless communication, and computer security. He won several awards, including the ISSIP 2020 Distinguished Recognition Award, ICC 2017 and IWCMC 2017 Best Paper Awards, the 2016 EECS Outstanding Research Award, and the 2009 NSF CAREER Award. He serves/served as an Associate Editor for several journals, including IEEE Transactions on Mobile Computing, IEEE Transactions on Wireless Communications, IEEE Network, and IEEE Transactions on Vehicular Technology. He also chaired/co-chaired many IEEE conference programs/symposia, including the 2017 INFOCOM Demo/Posters program, the 2016 IEEE GLOBECOM Mobile and Wireless Networks symposium, and many others. He served as a Distinguished Lecturer for the IEEE Communication Society for 2016 and 2017. He is a Senior Member of IEEE.}

~\\ {\footnotesize {\bf Abdurrahman Elmaghbub} received the BS degree from Oregon State University in 2019, and is currently pursuing his PhD degree in the School of Electrical Engineering and Computer Science at Oregon State University. His research interests are in the area of wireless communication and networking with a current focus on applying machine learning to wireless device classification.}

~\\ {\footnotesize {\bf Seifeddine Mejri} received the diploma of engineering from the Ecole Superieure des Communications de Tunis, Tunisia, in 2016, and the MS degree from the School of Electrical Engineering and Computer Science at Oregon State University in 2020. His research focus is on developing wireless RF signal detection and classification techniques using high-order signal statistics.}

\end{document}

%% file: commands.tex
\newcommand{\comment}[1]{ }

\newcounter{tc}

\newcommand{\myitemizebegin}{\begin{list}{$\bullet$}
{
 \setlength{\leftmargin}{0.4cm}
 \setlength{\parsep}{0.0cm}
 \setlength{\itemsep}{0.05cm}
 \setlength{\topsep}{0.0cm}
}}
\newcommand{\myitemizeend}{\end{list}}

\newcommand{\myitemizebeginless}{\begin{list}{$\bullet$}
{
 \setlength{\leftmargin}{0.7cm}
 \setlength{\parsep}{0.0cm}
 \setlength{\itemsep}{0.05cm}
 \setlength{\topsep}{0.0cm}
}}
\newcommand{\myitemizeendless}{\end{list}}

\theoremstyle{definition}

\theoremstyle{remark}

%% file: abstract.tex
Most prior works on deep learning-based wireless device classification using radio frequency (RF) data apply off-the-shelf deep neural network (DNN) models, which were matured mainly for domains like vision and language.
However, wireless RF data possesses unique characteristics that differentiate it from these other domains. For instance, RF data encompasses intermingled time and frequency features that are dictated by the underlying hardware and protocol configurations. In addition, wireless RF communication signals exhibit cyclostationarity due to repeated patterns (PHY pilots, frame prefixes, etc.) that these signals inherently contain.
In this paper, we begin by explaining and showing the unsuitability as well as limitations of existing DNN feature design approaches currently proposed to be used for wireless device classification. We then present novel feature design approaches that exploit the distinct structures of the RF communication signals and the spectrum emissions caused by transmitter hardware impairments to custom-make DNN models suitable for classifying wireless devices using RF signal data.
Our proposed DNN feature designs substantially improve classification robustness in terms of scalability, accuracy, signature anti-cloning, and insensitivity to environment perturbations. We end the paper by presenting other feature design strategies that have great potentials for providing further performance improvements of the DNN-based wireless device classification, and discuss the open research challenges related to these proposed strategies. 

%% file: introduction.tex
This paper is concerned with machine learning-based techniques that exploit radio frequency (RF) spectrum information to classify wireless devices. Although the study of the wireless classification problem already dates back to a few decades ago~\cite{weaver1969automatic}, much has changed over time. These changes include the methods and tools being used, the technological capabilities becoming available, and the increasing relevance and importance of the wireless classification problem due to several, newly emerging wireless applications such as spectrum access enforcement, dynamic spectrum sharing, and network access anomaly detection.
More specifically, the solutions are recently being shifted from conventional, model-based machine learning approaches using handcrafted features (e.g., Bayesian networks) to data-driven, function-based machine learning approaches capable of representation learning (e.g., deep neural networks (DNNs)). The application scenarios are also being shifted from detecting and identifying within a predefined small class space (e.g., few modulation types) to excessively large spaces, resulting from the growing diversity and scale of wireless devices due, for instance, to the large numbers of emerging IoT devices with their diverse hardware and protocol configurations. This shift is motivated and enabled primarily by recent advances in computing technology that are making data-driven approaches feasible in ways that were previously impossible.

Most previous works on wireless device classification that are RF data driven apply off-the-shelf deep neural network (DNN) models, which were primarily developed for the vision and language domains.
However, wireless RF data has characteristics and structures that are different from those of the vision and language domains. For instance, RF data embeds intermingled time and frequency features that are impacted by the configurations and information of the underlying hardware and protocol implementations. Another key difference is that wireless RF communication signals exhibit cyclostationarity because of the repeated patterns (e.g., PHY pilots, frame prefixes, etc.) that these signals inherently contain.
%
In recent years, research has also been focusing on exploiting signal distortions, caused by hardware impairments during manufacturing, to extract device features that can be used to increase classification accuracy. However, the accuracy achieved under these prior hardware-specific DNN approaches degrades with the decrease in the impairment variability among the devices, rendering such approaches unfit for classifying devices with high-end, minimally-distorted hardware components.

In this paper, we propose novel DNN feature designs that exploit the distinct structures of the RF communication signals as well as the hardware-impaired spectrum emissions to improve the robustness and efficiency of DNN-based device classification in terms of scalability, accuracy, signature anti-cloning, and insensitivity to environment perturbations.
More specifically, in this paper,
\myitemizebegin
\item We propose to expand the DNN feature set to include {\em out-of-band} spectrum emissions that are caused by hardware impairments to capture unique device signatures that efficiently separate among bit-similar, high-performance devices. This new feature set increases classification accuracy and signature anti-cloning substantially, yet without requiring any change in transmitters' hardware.

\item We exploit the cyclostationarity property of RF signals to extract signatures from signal patterns to improve learning latency and invariance to environment distortions.
    We propose to feed spectral correlation function (SCF) values of RF data as features to the DNN classifiers instead of the raw RF data, and we show that by doing so, the proposed feature designs reduce the sensitivity of the classifiers to channel and noise distortions substantially.
\item We propose and discuss other feature design strategies to further improve the classification performance.
    We show that by leveraging high-order data statistics and dimensionality compression techniques, we can further enhance the classification accuracy and learning overhead, as well as the robustness to channel impairments and system noise.
\myitemizeend

The paper is organized as follows. We start, in Section~\ref{sec:limiations-features}, by explaining the unsuitability and limitations of existing feature design approaches being currently used for DNN-based device classification. We end Section~\ref{sec:limiations-features} by presenting key goals and guidelines for designing efficient DNN features suitable for classifying wireless devices. We then propose, in Sections \ref{sec:hw-features} and \ref{sec:domain}, two complementary RF data-driven feature design approaches for efficient device classification.
Finally, in Section \ref{sec:open-challenges}, we present new feature design strategies that can provide further classification performance improvements, and also discuss the open challenges related to these proposed strategies.

%% file: prior-feature-limitation.tex
Previous RF data-driven DNN classification approaches mostly extract their features from hardware and/or protocol information, which provide radios with signatures that can distinguish them from one another. Protocol-specific features capture patterns that are extracted from repeated symbols such as PHY pilots and MAC headers, whereas hardware-specific features capture distortions in the signals that are caused by manufacturing errors in transmitters' hardware.

\subsection{Limitations of State-of-the-Art DNN Feature Designs}\label{subsec:lim}
DNN features in prior approaches are mostly extracted directly from raw RF data (i.e., IQ samples) of transmitted RF signals, without any prior processing or transformation of the input data. Though this makes their design simple, these feature designs suffer from the following major limitations:

\begin{figure}
	\center{
\includegraphics[width=\columnwidth]{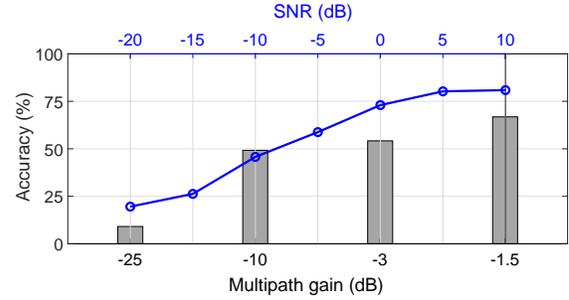}
	\vspace{-0.3in}
	\caption{Sensitivity to channel and noise impairments. Solid line (in blue): accuracy vs. SNR values; Bar graph (in black): accuracy vs. multipath gains}
    \label{fig:snr-gain-combined}}
\end{figure}

\subsubsection{Sensitivity to channel and noise impairments}
Features that are learned directly from raw IQ input data are proven to be sensitive to channel condition variation and noise. 
To illustrate, we simulated and measured the classification accuracy of DNN classifiers with features extracted from raw IQ data under varied SNR values and multipath gains. In this experiment, RF signals all communicated over a Rayleigh channel using eleven different modulation schemes (BPSK, 8PSK, 16QAM, 64QAM, FSK, CPFSK, GFSK, SSB-AM, DSB-AM, PAM4 and B-FM) are classified using a DNN classifier with raw IQ samples fed as its input features.
More details about experiment setup and dataset generation are provided in Section~\ref{subsec:domain-perf}.
Our results~\cite{seif-work} summarized in Fig.~\ref{fig:snr-gain-combined} show that the accuracy of the DNN classifier with raw IQ-based features exhibit high sensitivity to both noise and channel quality. For instance, the accuracy shrinks down from about 75\% under an SNR of 5dB to about 25\% under an SNR of -15dB. Likewise, we also observed that the accuracy shrinks from about 75\% under an multipath gain of -1.5 dB to about 10\% under a multipath gain of -25dB.

\subsubsection{Low separability among same-family protocols}
\begin{figure}
	\center{
\includegraphics[width=\columnwidth]{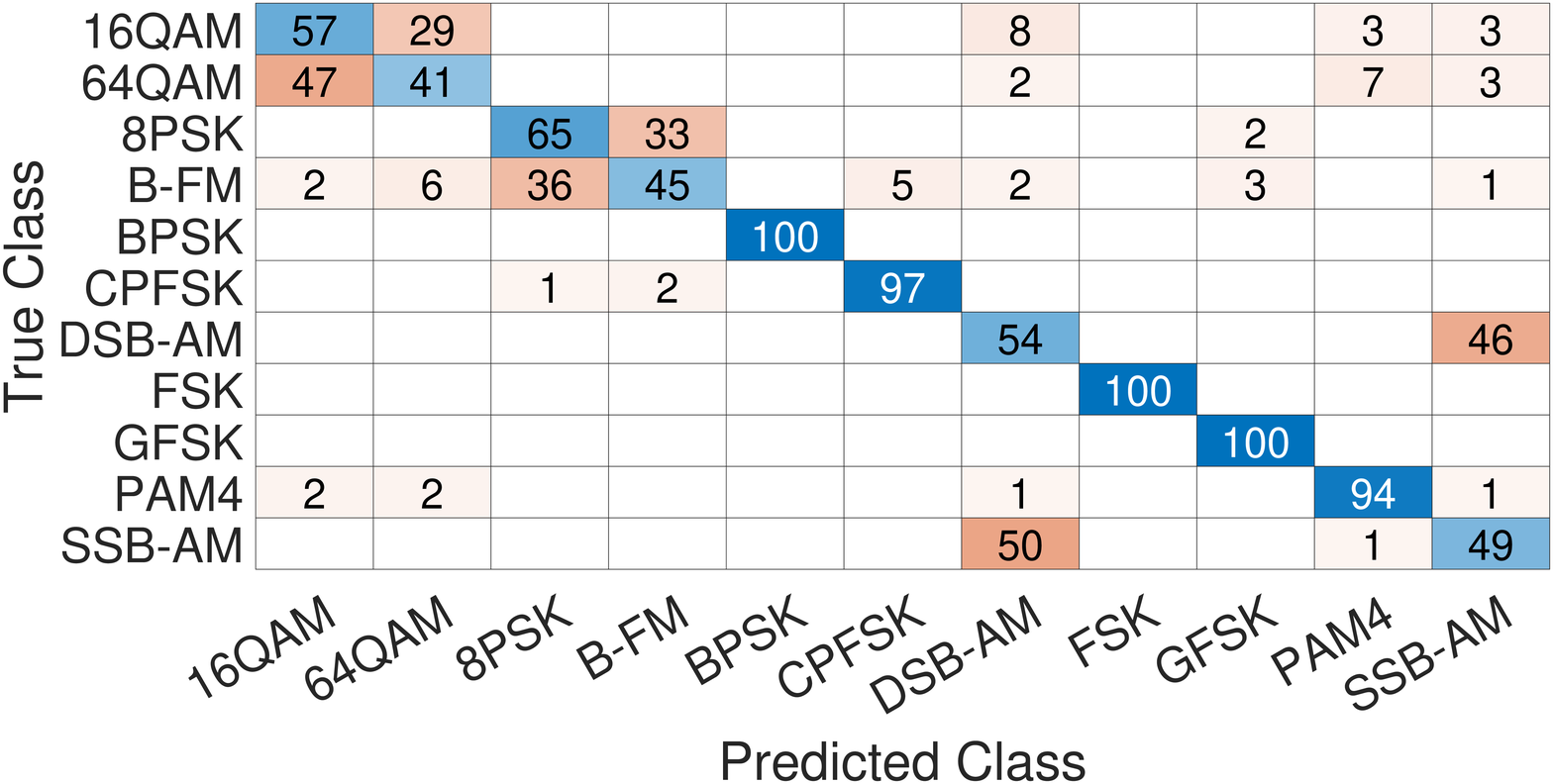}
	\caption{{Low separability}}
    \label{fig:IQ-SCF-1}}
\end{figure}

Features directly extracted from raw IQ data have also been proven to lead to low separability and longer training times.
In Fig.~\ref{fig:IQ-SCF-1}, we depict the confusion matrix of the IQ-based DNN classifier for RF signals using the eleven different modulations described above. Observe that the classifier fails to distinguish among schemes belonging to the same modulation family. For instance, there is a clear confusion between 16QAM and 64QAM and between SSB-AM and DSB-AM.
We show later that some basic transformation of the raw input data prior to feature extraction can not only improve class separability significantly, but also reduce training times by 5x without compromising the accuracy.

\subsubsection{Low separability among devices with bit-similar hardware}
Because hardware-specific features rely on the difference in the hardware impairment values to separate devices, their achievable classification accuracy decreases with the decrease in the impairment variability. In other words, when devices have similar impairment values, these features lead to poor classification accuracy. In addition, the signatures of these devices become more sensitive to distortions that arise from wireless channel and noise impairments. For instance, software-defined radios, such as USRP X310 radios, exhibit low impairment variability because they are made of high-performance hardware components, which make them not easy to classify using existing hardware-driven features.
Furthermore, as technology advances, the manufacturing impairment variations across devices are becoming extremely insignificant and the hardware is being made with lesser distortion. This also makes these prior classification approaches inaccurate and unscalable when applied to such devices.

\subsection{Feature Design Goals}

We propose feature designs that combine hardware impairments and protocol-specific information to:
\myitemizebegin
\item {\em Improve scalability} by maintaining high separability accuracy in the presence of massive numbers of devices with diverse protocol and hardware configurations.
\item {\em Ensure robustness against device signature cloning} by incorporating out-of-band TX hardware impairments into feature designs, making it too hard to recreate or replicate.
\item {\em Reduce sensitivity to environment distortions} by leveraging high-order statistics and deep learning to propose features that are agnostic to noise and channel variability.
\item {\em Reduce learning latency} by proposing hybrid features that combine model-based approaches with deep learning capability to accelerate learning tasks.

\myitemizeend

It is important to mention that even feeding raw IQ data as input to the DNN classifiers will eventually allow the DNN to learn to classify based on both hardware impairments and protocol-specific information.
However, the key challenge here---and hence the novelty of our techniques that we present in the next sections---lies in devising efficient strategies for providing 'denoised' extraction of these features, so as to achieve high separability and low learning latency while being agnostic to channel and noise distortions.

%% file: hw-features.tex
Unlike protocol-specific features, features that are extracted from hardware impairments are scalable and less susceptible to cloning and impersonation. As a result, recent works proposed to extract and use features from hardware errors and distortions incurred during manufacturing to improve classification performances.
As discussed in Section~\ref{sec:limiations-features}, these prior feature designs do not scale well and cannot provide high enough device separability accuracy when considering realistic environments---with varying wireless channel impairments---and newly emerging massive devices---with diverse protocol/hardware configurations and/or reduced hardware distortions.

\subsection{Leveraging Out-of-Band (OOB) Spectrum Distortion Information: A Novel Approach}
We now present a novel technique that substantially increases the performance of classifying bit-similar, high-performing transmitters. The proposed technique (i) is scalable in that it can distinguish among large numbers of devices made with minimally-distorted hardware,
(ii) is robust against signature cloning and replication,
(iii) requires no changes in transmitters' hardware, and
(iv) incurs minimal extra processing at the receiver side that can be performed with existing smart radios.

The novelty of our technique lies in considering both the {\em in-band} and {\em out-of-band (OOB)} spectrum emissions of received RF signals, caused by hardware impairments, to capture unique device signatures that efficiently discriminate among devices, even when devices are made with same, minimally-distorted hardware.
OOB emissions are those that predominate the out-of-band domain, defined as the frequency range separated from the assigned emission frequency by less than 250\% of the message bandwidth~\cite{tanaka2008unwanted}. OOB emissions are mainly caused by the modulation and the nonlinearity of the RF transceiver front-end, which result in an inevitable signal leakage into adjacent bands, despite the endless engineering and research efforts aimed at reducing it~\cite{morgan2006generalized}. Our proposed technique exploits such OOB emissions to provide unique device signatures and increase device separability and classification accuracy.

Due to manufacturing errors, the various hardware components of transmitters---including mixers, local oscillators (LO), and power amplifiers (PA)---are built with impairments. Such impairments manifest themselves in various signal and spectrum distortions---including phase noise, DC offset, IQ mismatch and others---that result in inevitable OOB emissions.
In this section, we take a closer look at the sources and impact of two key impairments, LO phase noise and PA nonlinearity distortion, caused by the transmitter's LO and PA components. Detailed analysis of these two impairments as well as other impairment manifestations (e.g., IQ mismatch, DC offset, quantization noise) can be found in~\cite{elmaghbub2020leveraging}.

\subsubsection{LO Phase Noise}
Typical direct-conversion transmitters leverage the quadrature mixer configuration to upconvert (separately and in parallel) the in-phase (I) and quadrature (Q) components of the baseband signal at the carrier frequency $w_c$ with two independent mixers fed by the LO tone shifted by 90$^o$ from one another.
That is, first, the in-phase signal component is multiplied (by one mixer) with the oscillating signal coming from the LO port and the quadrature signal component is multiplied (by the second mixer) also with the oscillating signal coming from the LO port but shifted by 90$^o$. Then, these two mixers' outputs are summed up to yield the upconverted signal modulated at the carrier frequency $w_c$.
Here, the LO is responsible for generating periodic oscillating signals that the mixers use to upconvert the baseband signal at the carrier frequency. That is, for an ideal LO, this periodic signal can be represented as a pure sinusoidal waveform $\cos(w_ct)$, which allows to upconvert baseband signals at the carrier frequency $w_c$ while preserving their original spectrum shape.

However, due to hardware impairments, the time domain instability of the oscillating signals generated by real LOs causes random phase fluctuations that result in expansion or regrowth of the signal spectrum in both sides of the carrier frequency. In other words, a real LO oscillating signal can be represented as $\cos(w_ct + \theta(t))$, where $\theta(t)$ represents the phase noise, which results in a random rotation of the signal constellation observed at the receiver and thus incurs undesired OOB emission.
%
To illustrate this OOB emission, consider applying the Fourier transform to the output of the in-phase mixer, $S_I(t)\cos(w_ct+ \theta(t))$, which is the product of the in-phase baseband signal, $S_I(t)$, and the real LO signal, $\cos(w_ct + \theta(t))$. Straightforward Fourier analysis shows that this phase noise term, $\theta(t)$, results in a bandwidth expansion or regrowth beyond the original signal's spectrum around the carrier frequency $w_c$. This bandwidth expansion originates from the convolution of the spectrum of the upconverted signal centred around $w_c$ and that of the phase noise (or precisely Fourier transform of $e^{-j\theta(t)}$). We refer the readers to Eq.~(7) of \cite{elmaghbub2020leveraging} for formal expressions of this Fourier analysis, illustrating the bandwidth expansion beyond the original signal's spectrum bandwidth.

\begin{figure}
	\center{
\includegraphics[width=\columnwidth]{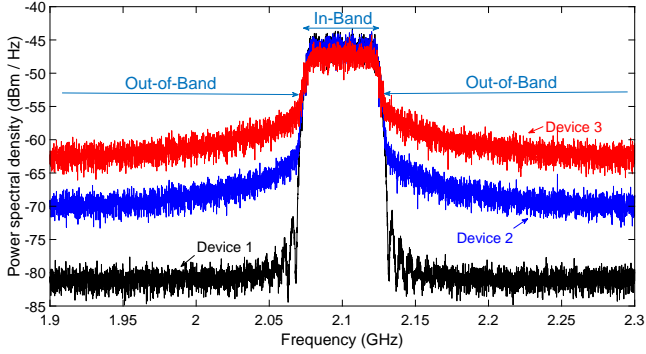}
	\caption{Phase Noise Effect: Device 1 (ideal LO); Device 2 (phase noise = -80 dBc/Hz); Device 3 (phase noise = -72 dBc/Hz); at 1MHz frequency offset.}
    \label{fig:Phase_noise}}
\end{figure}

Now since the spectrum regrowth depends on the LO's phase noise, different devices will exhibit different OOB distortions. This can be seen in Fig.~\ref{fig:Phase_noise}, which simulates three devices with different phase noise values but same frequency offset of 1MHz. Device $1$ mimics an ideal LO (i.e., zero phase noise), while Device $2$ and Device $3$ mimic real LOs with phase noise values of $-80$ and $-72$ dBc/Hz, respectively. The figure clearly shows that OOB spectrum shapes for Device $2$ and Device $3$ are different from one another and from Device $1$.
As will be shown in Section~\ref{subsec:hw-results}, our technique exploits these OOB distortions, caused by the transmitters' phase noises as well as other hardware impairments, to enhance device classification performance. 

\subsubsection{PA Nonlinearity Distortion}
\begin{figure}
	\center{
\includegraphics[width=\columnwidth]{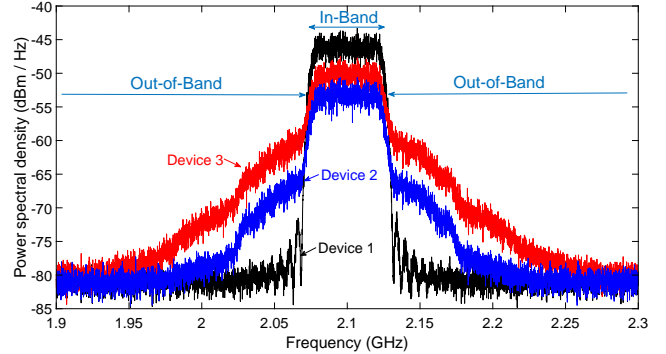}
	\caption{Nonlinearity effect under 16QAM modulation}
    \label{fig:PA}}
\end{figure}

The majority of circuit nonlinearity is attributed to the PA whose main job is to boost the modulated RF signal with enough radiation power to allow it to reach its target destination.
The PA's nonlinear output in response to an input signal, ${S}(t)$, can be modelled as~\cite{gard2005impact}
$${\alpha}_1 {S}(t) + {\alpha}_3 {S}^3(t) +  {\alpha}_5 {S}^5(t) + ...$$
where ${\alpha}_i$s are model parameters.
To understand the impact of PA nonlinearity on OOB spectrum distortions, suppose the PA input signal is $S(t)=A(t)\cos(w_ct+\phi(t))$ and consider illustrating the effect of the third-order nonlinear term, ${\alpha}_3S^3(t)$, only, which can be written as
\begin{equation}\nonumber
    \frac{{\alpha}_3A^3(t)}{4}\big[3\cos(w_ct+\phi(t)) + \cos(3w_ct+3\phi(t))\big].
\end{equation}
Now provided that the out-of-band component at $3w_c$ is located sufficiently far away from the center frequency, $w_c$, and that the bandwidth of the original signal is much less than $w_c$, this out-of-band component can easily be filtered out without causing any bandwidth regrowth around the original message spectrum. However, the first term at $w_c$ can lead to spectrum regrowth.
For instance, in the case of constant-envelope modulation schemes such as BPSK where the amplitude $A(t)$ is constant, the spectrum of the modulated signal in the vicinity of $w_c$ clearly remains unchanged.
However, for variable-envelope modulation schemes such as $16$QAM where the amplitude $A(t)$ varies over time, nonlinearity causes a spectral regrowth of the original signal spectrum because the ${{\alpha}_3A^3(t)}/{4}$ term generally exhibits a broader spectrum than $A(t)$ itself.
For this case of modulation, the severity of the spectral growth also depends on the nonlinearity model parameter ${\alpha}_3$.
To illustrate, we show in Fig.~\ref{fig:PA} the case of a 16QAM modulated signal passing through a linear PA (Device 1) and two nonlinear PAs (Devices 2 and 3) each under slightly different nonlinearity parameters.
Two key observations we make from these results. First, observe that the nonlinearity of PA leads to an OOB spectrum growth (or distortion). Second, even a slight difference in the PA nonlinearity impairments causes considerable differences in the amplitude of the frequency components in the OOB spectrum.
Our proposed technique exploits such an OOB distortion information to increase device distinguishability, thereby enhancing device classification performance. More details on this proposed OOB technique can be found in~\cite{elmaghbub2020leveraging,elmaghbub2020exploiting}.

\subsection{Performance Evaluation}\label{subsec:hw-results}
We used MATLAB's Deep Learning Toolbox, running on a system with Intel Core i7 8th Gen CPU, to implement a convolution neural network (CNN) model, with input feature being the collected IQ samples, each represented as a $2\times1024$ real-valued tensor.
We chose to use CNN because it is proven to be an effective model for wireless classification (see~\cite{o2016convolutional}).
The first convolutional layer has $16$ $1\times4$ filters, with each filter learning $4$-sample variations in time over the I or Q dimension.
After each convolutional layer follows a Batch normalization layer, a ReLU activation, and a Maximum Pooling layer with filters of size $1\times2$ and stride [$1$ $2$]. The last convolutional layer is followed by an Average Pooling layer with a dimension $1\times32$, which is then given as an input to the Fully Connected layer whose output is finally passed to a softmax classifier layer.
We set dropout rate to $0.5$ to address overfitting, and use the gradient descent with momentum
for weight optimization. Error is minimized via back-propagation with categorical cross-entropy used for loss function. Refer to~\cite{elmaghbub2020leveraging} for additional setup detail.

\begin{figure}
\centerline{
    \subfigure[Proposed technique.]
    {\includegraphics[width=.5\columnwidth,height=0.22\textwidth]{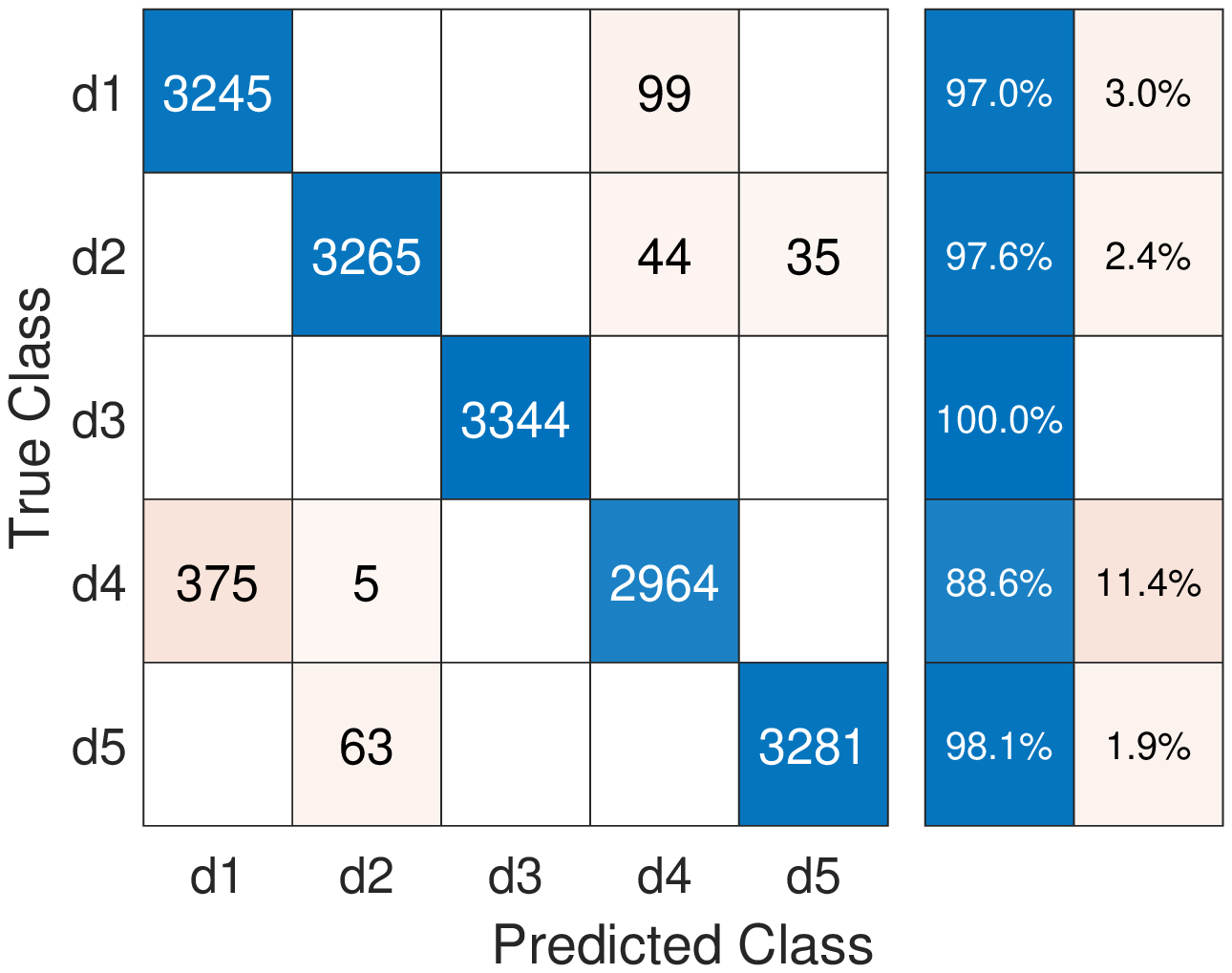}
    \label{subfig:conf-mat-f1}}
\hspace{-0.15in}
    \subfigure[In-band only.]
    {\includegraphics[width=.5\columnwidth,height=0.22\textwidth]{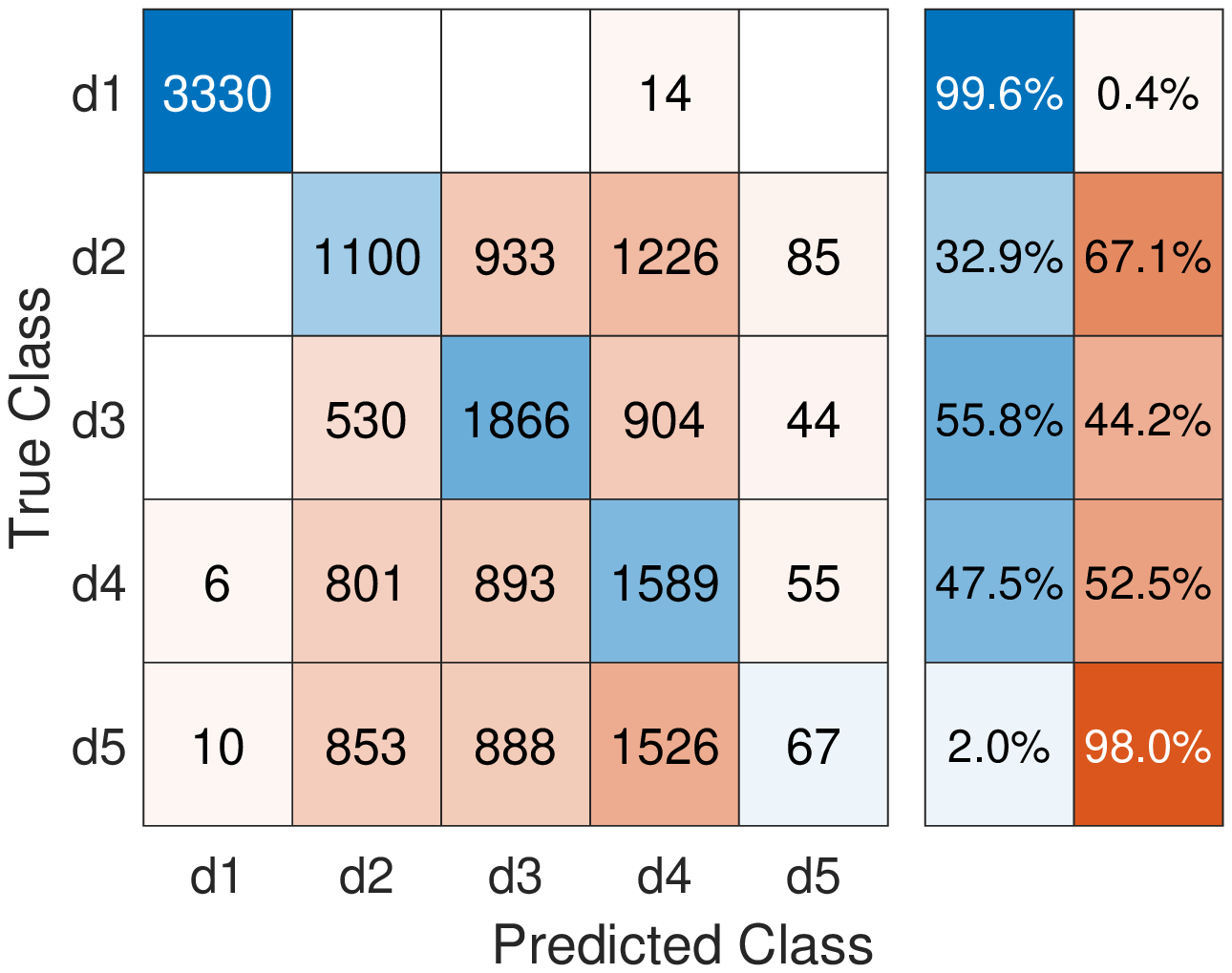}
    \label{subfig:conf-mat-f2}}
    }
\caption{{(a) proposed technique (both out-of-band and in-band); (b) conventional technique (in-band only).  d$_i$ indicates Device $i$.}}
\label{fig:conf-mat-f}
\end{figure}

We emulated five hardware-impaired devices to compare the performance of the proposed technique, leveraging both in-band and out-of-band spectrum distortion information, and that of the conventional technique, using in-band distortion information only. The impairment values of the devices are set very similar to resemble bit-similar, high-performing radios. Each device sends $16$QAM modulated signals over an AWGN channel with IQ values being collected for two different bandwidths, $2.075$ -- $2.125$ GHz, which represents the bandwidth of the message (in-band), and $1.9$ -- $2.3$ GHz, which includes both in-band (message bandwidth) and out-of-band spectra.
%
Our results, depicted in Fig.~\ref{fig:conf-mat-f}, show that our technique achieves substantially higher accuracy than that of the in-band only technique, with a testing accuracy of about \textbf{96\%} under our technique versus only about \textbf{50\%} under the in-band only technique. Our results show that exploiting OOB distortion information caused by radio hardware impairments increases device separability substantially.

%% file: domain-features.tex
We now present feature designs that
build on the well-established cyclostationary analysis theory~\cite{gardner1986statistical} to complement the hardware-specific features presented in Section~\ref{sec:hw-features}. This theory exploits the fact that wireless communication signals exhibit cyclostationarity~\cite{gardner1986statistical}, because they inherently contain periodic patterns due, for instance, to repeated PHY pilots, symbol prefixes, MAC headers, and so on. This means that although the raw signal itself does not implicitly contain repeating patterns, high-order statistics (moments and cumulants) of the data exhibit periodic patterns that can be exploited and extracted to serve as unique signal features.

\subsection{Leveraging Second-Order Data Statistics: A Spectral Correlation Function (SCF) Based Approach}\label{subsec:scf}
Provided that the signal received and sampled by the receiver is cyclostationary, its autocorrelation function, $R(t,\tau)$, is periodic in time $t$ and peaks when the delay lag $\tau$ corresponds to the period at which the pattern repeats.
In other words, the autocorrelation function can be expressed as a Fourier series,
$$\sum_{\alpha}{R^{\alpha}{(\tau)}}e^{j2\pi \alpha t},$$
with a sum taken over cyclic frequencies $\alpha$ that are integer multiples of the fundamental frequency of the repeated pattern, and the Fourier coefficient ${R^{\alpha}{(\tau)}}$ peaks only at cyclic frequencies $\alpha$ and delay lags $\tau$ that correspond to the repeating pattern of the signal. Now because if the pattern repeats every $\tau$ then it will repeat every integer multiple of $\tau$, the Fourier transform of the autocorrelation function, also known as the spectral correlation function (SCF),
will peak at angular frequency $f=1/\tau$ if the Fourier coefficient ${R^{\alpha}{(\tau)}}$ peaks at delay lag $\tau$. That is, SCF, which is a function of both angular and cyclic frequencies $f$ and $\alpha$, will peak at frequencies $f$ that are inversely proportional to the delay lags $\tau$ that correspond to the repeated pattern.
Therefore, the frequencies of the repeating pattern embedded in the signal can be used to increase separability across different PHY signals, and hence, can serve as protocol-specific features.

\subsection{Performance Evaluation}\label{subsec:domain-perf}
We used MATLAB's Communication and Deep Learning Toolboxes to generate a dataset for eleven modulated signal types, BPSK, QPSK, 8-PSK, 16-QAM, 64-QAM, PAM4, GFSK, CPFSK, B-FM, DSB-AM, and SSB-AM, using a Rayleigh channel with multipath gains: $-25$, $-10$, $-3$, $-1.5$, $0$ dB. The dataset contains $1000$ frames/samples for each modulation type.
For the DNN model, we used a CNN architecture similar to the one presented in Section~\ref{subsec:hw-results}, consisting of 2D convolutional layers with $8$ and $16$ features maps, respectively, two Batch Normalization layers, and one fully connected layer with output being fed to an $11$-way softmax that produces a distribution over the $11$ modulation classes.
Additional detail on this setup can be found in~\cite{seif-work}.

\begin{figure}
    \centering
    \includegraphics[width=\columnwidth]{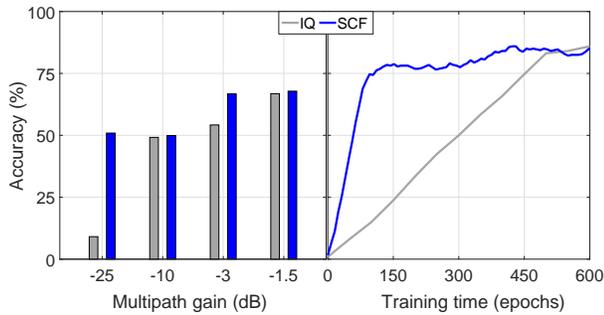}
    \caption{Comparison of classification accuracy and training time performances: SCF-based vs. IQ-based features.}
    \label{fig:scf-iq-training}
\end{figure}

In Fig.~\ref{fig:scf-iq-training}, we compare the classification performance obtained under the proposed approach, using SCF as the DNN features, and the existing approach, using IQ samples as the DNN features. The (right) figure shows that SCF-based features can reduce learning time by 5x, without impacting the accuracy (both achieve about 80\% accuracy for this experiment). In addition, the (left) figure shows that these SCF-based features are less sensitive to channel condition variations caused by varied multipath gains.
Although high-order statistics based features (e.g., SCF) show promising results, there remains key challenges that need to be addressed, some of which are discussed in Section~\ref{sec:open-challenges}.

%% file: open-challenges.tex
We now present new feature design ideas and strategies that provide further improvements of the classification performance, with again a focus on accuracy, learning overhead, and invariance to channel impairments and system noise.

\subsection{Cyclic Frequency Estimation for Nonblind Classification}
%
For the SCF-based DNN classifiers proposed in Section~\ref{sec:domain}, having prior knowledge of the cyclic frequencies (CFs) can improve the achievable classification performance substantially.
To illustrate this, we performed an experiment for classifying PHY signals with the eleven different modulation schemes studied and described in Section~\ref{subsec:lim}, where analytically-derived CFs~\cite{kim2007cyclostationary} (one for each modulation type) are used for nonblind classification, and random CFs are used for blind classification. Our results~\cite{seif-work} show that nonblind (i.e., with known CFs) classifiers achieve an overall classification accuracy of about 70\% whereas blind (i.e., with unknown CFs) classifiers achieve an accuracy of about 45\% only.
Several well-established techniques (e.g., TSM and SSCA~\cite{spooner_tunneling,brown1993digital}) for estimating the CFs and SCF have already been developed over a couple of decays worth of research. These techniques, however, are computationally complex, making them unsuitable for real-time application scenarios.

\comment{
\begin{figure}
\center{
    \subfigure[\scriptsize Non-blind classifier]
    {\includegraphics[width=.8\columnwidth,height=0.2\textwidth]{CM-SCF-SNR-5-nonblind}}
    \label{subfig:c1}}
    %
\center{
    \subfigure[\scriptsize Blind classifier]
    {\includegraphics[width=.8\columnwidth,height=0.2\textwidth]{CM-SCF-SNR-5-blind}}
    \label{subfig:c3}}
    %
\caption{{Blind/nonblind}}
\label{fig:cf}
\end{figure}
}

One potential solution is to investigate the use of SSCA, a computationally efficient method for cyclic feature analysis of second-order cyclostationary communication signals with unknown CFs~\cite{brown1993digital}.
SSCA takes as input IQ data, sampling frequency, spectral resolution and  cyclic resolution, and outputs CFs and SCF.
One approach is to first use SSCA to compute (offline) CFs for different IQ input samples, and then use these computed CFs along with their corresponding IQ samples to train DNN whose output would be the CFs. During the inference phase, this trained SSCA DNN is to be used for real-time computation of the CFs, which are then fed (along with the other features) to the DNN classifier for device classification.

\subsection{Exploiting High-Order Data Statistics}
So far, we have shown that using order-2 cumulants (e.g., SCF-based) as features improve classification performance. Another potential idea is to consider using high-order cumulants; for formal definitions of cumulants, we refer readers to~\cite{gardner1994cumulant-1}. To illustrate, we run an experiment where order-2 and order-4 cumulants are used as features for the DNN classifier to separate between different RF signals modulated with QPSK, 16QAM and 64QAM, and our results (not shown in this paper; see \cite{seif-work} for details) reveal that order-4 cumulant based features offer better separability than using order-2 cumulants only. Another benefits of designing features based on high-order cumulants is that they have been proven to be less sensitive to noise~\cite{spooner2017modulation}. Though promise much better separability, they are, however, also much harder to compute and hence pose some feasibility challenges. One approach, which requires further investigation, for overcoming this challenge is to consider hybrid model-based and data-driven DNN strategies that allow to extract features from these complex models through deep learning, and then use these features for device classification.

\subsection{Feature Compression and Dimensionality Reduction}
The performance of the SCF-based approach presented in Section~\ref{subsec:scf} depends heavily on the (cyclic and angular) frequency resolutions chosen during SCF sampling. This is because both accuracy and training time depend on the SCF input size. More specifically, with SCF, the higher frequency resolution, the higher the accuracy, but also the longer the training time. One potential approach for addressing this issue is {\em compressed learning}, which provides a projection from the data domain to a measurement domain that preserves the linear separability under certain conditions \cite{calderbank2009compressed}. Classification in the measurement domain can still be possible (without compromising accuracy) after transforming the data to some appropriate compressed domain, provided that the linear projection preserves the structure of the instance space.
Compressed learning can be exploited to reduce SCF's high dimensionality, thus reducing its computational complexity during training time. However, although proven effective in many other classification applications like image recognition, natural language processing, sensor networks, and automatic modulation recognition, much work remains to be done to apply compressed learning for wireless device classification.

%% file: conclusion.tex
This paper presents new DNN feature designs that rely on RF data to classify wireless devices. The proposed designs exploit the distinct structures of RF communication signals and the hardware impairments incurred during device manufacturing to custom-make DNN models to improve classification scalability and accuracy, signature anti-cloning, and insensitivity to environment distortions. The paper also explains the limitations of existing DNN features when applied to device classification, and presents other feature design ideas and discusses their related challenges. 